\documentclass{aastex}
\usepackage{emulateapj5}

\shorttitle{Strong lensing of supernovae}
\shortauthors{Holz}

\begin{document}

\title{Seeing double: strong gravitational lensing of
high-redshift supernovae}
\author{Daniel E. Holz}
\affil{Institute for Theoretical Physics, University of
California, Santa Barbara, CA 93106}

\begin{abstract}
With the advent of large, deep surveys, the observation of a
strongly gravitationally lensed supernova becomes
increasingly likely.  High-redshift surveys continue apace,
with a handful of type Ia supernovae observed to date at
redshifts of one or greater. In addition, a satellite (the
{\em Supernova/Acceleration Probe}\/[SNAP]) has been
proposed dedicated to observing thousands of supernovae per
year out to a redshift of $1.7$. Although it is exceedingly
unlikely that we will see a multiply-imaged supernova from
ongoing surveys, we find that SNAP would observe at least eight
such events per year. Since having a standard candle is
inessential to most lensing studies, SNAP's large sample of
type II supernovae contributes to this rate. Each case of
strong lensing allows for a precise determination of time
delays, image separations, and relative image
magnifications, and the SNAP strong-lensing database will
offer measures of $\Omega_m$, $\Omega_\Lambda$, and $H_0$,
{\em independent} of SNAP's primary goal of establishing the
distance-redshift relation.  These systems also constrain
models for the matter density profiles of galaxies and
clusters. Furthermore, lensed type Ia supernovae afford the
opportunity to break the mass-sheet degeneracy found in many
lensing measurements.
\end{abstract}


\section{Introduction}

Type Ia supernovae (SNe) have become an extremely powerful
cosmological probe. They appear to be very good (calibrated)
standard candles~\citep{rpk96a,branch98}, and because of
this they enable a direct measurement of the
distance-redshift relation, and hence the expansion history
of the universe~\citep{stir79,gp95}. At present on the order
of 100 high-redshift SNe have been observed, and surveys are
adding dozens of type Ia's a year at increasingly high
redshifts.  Results from these surveys indicate that the
universe's expansion is accelerating~\citep{riess98,perl99},
and this discovery has led to a tremendous amount of
excitement in the cosmological community.  In particular,
there is great interest in the SuperNova/Acceleration
Probe (SNAP), a satellite dedicated to searching for and
observing high-$z$ SNe.\footnote{see {\tt
http://snap.lbl.gov} for more information.} The primary
objective of the mission would be to measure cosmological
parameters to unprecedented accuracy, enabling, for example,
the distinction between a cosmological constant and a
``quintessence'' component~\citep{ht99}. The satellite would
achieve this by observing over 2000 type Ia SNe per year,
getting a complete light curve, multiple colors, and a
spectrum at maximum light for each one, and thereby
measuring the distance-redshift relation out to $z=1.7$ to
unprecedented precision.

Gravitational lensing can influence the observed brightness
of high-$z$ SN data. The impact of small shifts in the
brightnesses of SNe due to weak lensing has been extensively
explored, both as a potential source of
noise~\citep{kvb95,frieman97,wcxo97,holz98}, and as a source
of science~\citep{ms99,holz99,sh99,ben99}. In this paper we
focus on the possibility for strong lensing of SNe, leading
to multiple images of a given source.  While this
possibility has also been
discussed~\citep{sw87,lsw88,kp88,pm00}, we consider strong
lensing in the age of large-scale, high-$z$ SN surveys~(see
also~\citet{wang00}). Although current surveys are unlikely to
produce a case of multiple imaging, we find that SNAP would
be expected to see numerous such events. In addition, SNAP
would be well suited to do extensive characterization of the
events, bringing the very real possibility of science from
multiply-imaged SNe in the foreseeable future.

\section{Likelihood of multiply-imaged supernovae}

\label{S:likelihood}

We commence by estimating the likelihood of strong lensing.
An empirical value comes from the rate of observed
strong-lensing in optical quasar or radio source surveys.
In an attempt to control selection effects, the Jodrell-Bank
VLA Astrometric Survey (JVAS) and the Cosmic Lens All-Sky
Survey (CLASS) have conducted a statistically complete
survey of over $15,000$ flat-spectrum radio
sources~\citep{bm00,myers01}. With analysis almost complete,
they have found $18$ lensed sources, with $14$ having source
redshifts and image time delays that fall within the SNAP
detectability range. Taken at face value (e.g. neglecting
selection effects and population evolution), this would
indicate that approximately $0.1\%$ of SNe seen by SNAP
would be expected to be strongly lensed, with image separations on
the order of one arcsec and time delays between images
ranging from a few days to over a year.
As this statistic is based on a small 
number of strong lensing events, it is also instructive
to look at theoretical calculations of the lensing rate.
With this in mind we utilize the ``stochastic universe
method'' (SUM) to determine the probability for
multiple imaging~\citep{hw98}.
A key feature of this approach is that it accommodates
lensing from {\em all}\/ matter along the line of
sight---possibly including the effects from thousands of
galaxies---and in addition requires no assumptions
regarding the luminosity-distance relation
\citep{hw98,bggm00}. It is to be emphasized,
however, that the strong lensing results do not depend
sensitively on the details of the methods, with differing
analytic and numerical results being in good
agreement~\citep{tog84,hmq99,ms99}.

In Fig.~\ref{opt_depth1} we show the probability of
multiple-imaging of a given source (also referred to as the
optical depth to strong lensing, $\tau$) as a function of
the source redshift, for a range of cosmologies.  We take a
conservative value of the dimensionless lensing efficacy
parameter, $F=0.025$~(Turner et al. 1984).  The results are
relatively insensitive to details of the mass function, and
calculations with a PS distribution~\citep{ps74} of galaxy
masses yields similar results. Although taking all of the
matter in the Universe to be in isothermal spheres is a
great oversimplification, it is to be emphasized that
effective modeling of many strong-lensing systems is
accomplished with isothermal mass
distributions~\citep{kf99,wmk00,cohn01}.\footnote{Singular
isothermal spheres always produce double images, while
roughly half of observed strong lensing systems are
quads~\citep{rt01}. Adding background shear, or considering
singular isothermal ellipsoid lenses, produces the requisite
quad images. Such generalizations can be expected to raise
the optical depth curves in
Fig.~\ref{opt_depth1}~\citep{sef92}.} Although NFW
profiles~\citep{nfw96} are expected to be a better
approximation to the mass distribution of massive halos,
baryons are expected to isothermalize the matter
distributions for halos of galaxy mass and
below~\citep{kochwhite01}.
\citet{pm00} have undertaken a study of lensing in cold dark 
matter universes with a PS distribution of galaxy masses,
where NFW profiles are utilized for massive halos
($m>3.5\times10^{13}\mbox{ M}_\sun$), and singular
isothermal spherical profiles are used otherwise. Our
extremely simplified model agrees with their results to
within $\sim20\%$. Our results are also consistent with the
CLASS data.

Data in cosmology in recent years has been pointing to a
``cosmic concordance'' model~\citep{bops99}, with
$\Omega_m=1/3$, $\Omega_\Lambda=2/3$, and $H_0=70\mbox{
km}\,\mbox{sec}^{-1}\,\mbox{Mpc}^{-1}$, and we restrict our
attention to this model in what follows. The likelihood
results in Fig.~\ref{opt_depth1} imply that $0.05\%$ of
sources at redshift $z=1$ would be expected to be multiply
imaged on arcsecond scales. At present roughly $100$
high-$z$ SNe have been observed (mostly at $z\approx0.5$),
and thus it is quite unlikely that any of them are multiply
imaged. The situation is more bleak than the numbers
indicate, however, as even if one of the observed high-$z$
SNe happens to be multiply-imaged, it is exceedingly
unlikely that we would stumble upon successive images.

The situation is much brighter if SNAP flies.  SNAP's
``optical'' ($0.35$--$1.0\,\mu\mbox{m}$) imager would observe
3,800 type Ia SNe per year, with detailed follow-up
(restframe $B$-band photometry and spectra) of a subsample
of 2,400 (100 at $z>1.2$, and all at $z\leq1.2$). By
checking whether prior 
images have appeared nearby (within
$15\arcsec$) when selecting high-$z$ SNe for follow-up, SNAP
can assure that every multiply-imaged type Ia SN has at
least one
\resizebox{8.4cm}{!}{
\includegraphics{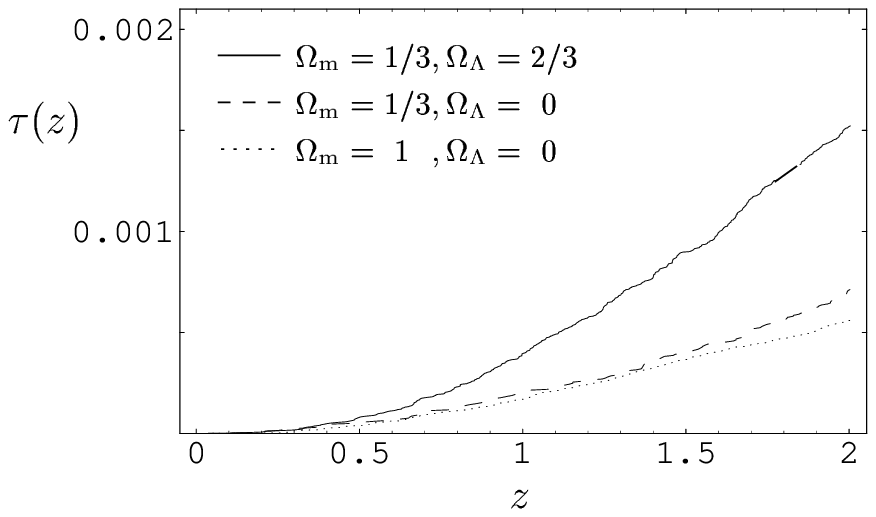}}
\figcaption[optical_depth7.eps]{The optical depth, $\tau$, to multiple imaging, as
a function of redshift, $z$. Results for a range of
cosmological parameters are shown, for a lensing efficacy of
$F=0.025$.
\label{opt_depth1}}
\vspace{0.25cm}
\noindent
image with a redshift and absolute magnification
determination. The ratio of the optical fluxes then allows a
determination of the absolute magnification of all other
images.
Convolving the optical depth curve of
Fig.~\ref{opt_depth1} with the expected redshift
distribution of SNAP SNe (relatively flat, with $1,500$ type
Ia's$/$year at $1.2<z<1.7$) yields an expectation of 2
multiply imaged Ia SNe per year.

SNAP would provide a homogeneously selected 28.5 (restframe
$U$ or $B$-band) magnitude-limited sample of all SNe which
occur during its observation period in its 20 fixed
$1\,\deg^2$ fields of view, with two fields being
additionally imaged down to a limiting magnitude of
30. Since the gap between deep images of each field will
always be less than eight days, no SNe will fall through the
observing cracks.  Although the
SNAP design is still in flux, we will take as one of its
core requirements that all type Ia SNe with $z<1.7$ will be
discovered at least 3.8 magnitudes below peak brightness
(and thus within $\sim2$ rest frame days of the
explosion). Therefore lensed images which have $>3\%$ of the
(unlensed) source flux will be observed. From our simplified
SUM calculations, we find that 94\% of strongly-lensed
events are at least this bright. SNAP will see every
type Ia SN out to redshift 1.7 in its deep fields, and will
observe the vast majority of strong lensing events of these
sources, thereby minimizing magnification bias in the strong
lensing sample.

SNAP will also see a very large sample of type II SNe
(though only a tiny fraction of these will have detailed
follow-ups).  While these are significantly dimmer (peak
value 2.3 magnitudes below the Ia peak (in restframe
$B$-band), with a dispersion of 1.3 mag~\citep{gnp99}),
they also occur at much higher rates (3 times more frequent
at present day, rising to 5--10 times more frequent at
redshift 1--2~\citep{sensm00}). Taking a conservative
value of 5 times the Ia rate, we find that SNAP would have
10 multiply-imaged type II SNe in its sample per year. SNAP
will catch the average SN II event only 1.5 magnitudes below
peak, and so will miss images demagnified to less than 25\%
of the unlensed flux. Convolving
the large intrinsic dispersion in type II peak brightness
with the distribution of image pair magnifications (from our
SUM calculations), we find that 60\% of multiply-imaged type
IIs will be observable, yielding an expected total of 6
events per year.  For these systems, SNAP's optical imager
light curve will provide a precise determination of the
image time delays, separations, and relative fluxes. As one
does not require the standard candle luminosities of lensed
SNe to do strong-lensing science (see Sec.~\ref{S:cosmo}
and~\ref{S:hubble}), strongly lensed type II systems are a
valuable contribution to the lens database.

Combining the rate for SNe Ia and II, we conservatively
estimate that SNAP would detect eight multiply-imaged SNe per
year. Let us therefore assume that SNAP will fly, and
explore the science that can be done with multiply-imaged SN
systems.

\section{Detection criteria}
\label{S:detectability}

A multiply-imaged SN would appear as successive SNe, closely
clustered on the sky ($\la 15\arcsec$). As gravitational
lensing is achromatic, the light curves of the ``different''
SNe would be identical in all bands, modulo an overall
amplification shift between the different
images.\footnote{Extinction along different lines of sight
could yield a chromatic difference between multiple images
of a given SN. However, because of the extensive
phenomenological knowledge of type Ia SN light curves and
colors~\citep{rpk96a,rpk96b}, the extinction in a given
image is quantifiable to within $\la0.04\mbox{ mag}$,
sharply reducing the likelihood of false
dissociations. Having multiple images sampling different
lines of sight through a galaxy or cluster could provide a
rudimentary map of the distribution of dust in the lens.}
The probability of having a second SN appear in a given
galaxy within one year is roughly equivalent to the
multiple-imaging rate, and therefore well-sampled photometry
of the light curves (to enable comparison of rise/fall
parameters), or searches for lenses along the line
of sight, will be required to positively identify lensing
cases.

Characteristics of the SNAP mission impose a number of
selection constraints on observable multiply-imaged SNe
systems.  The distribution of time delays between SN images
can be expected to mimic those in multiply-imaged radio
source and quasar surveys. From the theoretical
distributions of time delays obtained via the SUM
calculations, we find that $78\%$ of strong lensing events
have time delays such that they'd be seen in one year's
observation, rising to $93\%$ with three years of
observation.

The angular resolution of SNAP is to be
$0\farcs1$, which allows it to directly resolve the
vast majority of image separations.
An important
possible systematic error for ground-based optical surveys
of multiply-imaged systems is that images appearing near the
center of the lens can be obscured by the
lens~\citep{pm00}. Because of SNAP's excellent angular
resolution, and because baseline images of the galaxy
without the supernovae will be available (and hence can be
subtracted off), a SN image would be identifiable all the
way into the core of a galaxy.

Even if the angular separation of SN images is well below
the resolution limit of the SNAP telescope (which is
exceedingly unlikely for macrolensing cases), because the
SNe are transitory we may nonetheless be able to see and
distinguish images.  As long as the light curves do not
significantly overlap in time, proximity (and blending) of
images is no impediment to observation. In addition, it is
straightforward to associate even very widely spaced images,
as we expect to see all SNe within SNAP's large
($1\,\deg^2$) field of view (limited to systems with time
delays which fall within the SNAP observational period).

The likelihood results are quite sensitive to the form of
the dark matter. Thus far we have been considering smoothly
distributed isothermal halos filled with microscopic dark
matter (e.g. axions, WIMPS). If a significant ($\ga\!\!10\%$)
fraction of halo matter is in macroscopic form (e.g. MACHOs,
black holes), the lensing probability increases dramatically
due to frequent microlensing (Schneider \& Wagoner 1987;
Rauch 1991). At
cosmological distances, lensing by compact
objects with masses below $10^6\,M_\odot$ produces image
separations of less than a few milliarcseconds and time
delays of less than a minute, indicating that
multiple imaging would no longer be directly
discernible. The overall amplification of the combined image
would still be noticeable down to lens mass scales of
$\sim10^{-4}\,M_\odot$, below which the SN can no longer be treated
as a point source (linear extent $\sim 10^{15}\,\mbox{cm}$
corresponds to an angular size on the order of the Einstein
angle of the lens), and lensing ceases to have an effect. In
addition, if there is relative motion between the lens and
the line of sight to the source, the amplification
becomes time dependent.  The lensing timescale at
cosmological distances goes as $\delta t\sim18\
\mbox{days}\,(10^9\mbox{
cm}\,\mbox{sec}^{-1}/v)\sqrt(M/10^{-3}\,M_\odot)$,
with $v$ the expansion velocity of the SN, and $m$ the
mass of the compact lens. Therefore
lenses in the mass range $10^{-4}\,M_\odot\la m \la
10^{-3}\,M_\odot$ can produce a MACHO-type signal superposed on
the SN light curve. Microlensing by compact
objects with $m\ga10^{-3}\,M_\odot$ will manifest itself as an
overall amplification shift of the (standard candle) light
curve, inducing further scatter in the SN Hubble
diagram (Minty, Heavens, \& Hawkins 2001).

\section{Independent measure of cosmological constitution}
\label{S:cosmo}

Over the course of a few years, SNAP would provide a
uniformly selected survey of many thousands of SNe, offering
the possibility of a variety of statistical tests of
cosmology.  For example, the different curves in
Fig.~\ref{opt_depth1} should be distinguishable
observationally---three years' data would differentiate
between $\Omega_\Lambda=2/3$ and $\Omega_\Lambda=0$ at
better than $3\sigma$. That the SNe are standard candles is
completely irrelevant to determining cosmology via lensing
likelihood, and thus strong lensing provides a powerful
independent probe, and consistency check, for the SN
measurement of the distance-redshift relation (the primary
mission of SNAP).

In addition, the strong lensing likelihood depends upon the
clumpiness of the dark matter, and thereby allows the
distinction between microscopic and macroscopic dark
matter~(Linder et al. 1988). Furthermore, Wyithe et
al. (2001) have suggested that the optical depth to multiple
imaging could be used to distinguish between interacting and
non-interacting dark matter. A
complementary cosmological measurement is offered by the
distribution of image separations and lensing image
morphologies~\citep{kks97,rm01,kochwhite01}. These
approaches are less sensitive to the cosmological
parameters, but quite sensitive to the mass distribution
within halos. 

\section{Independent measure of cosmological scale}
\label{S:hubble}

Multiply-imaged systems allow for the determination of the
Hubble constant, $H_0$, through the measurement of time
delays between images~\citep{refsdal64}. This method has
been developed for over a decade in multiply-imaged quasar
systems, with varying success~\citep{kf99}. One of the
limitations of such systems is that the time delays can
often be quite difficult to determine~\citep{kundic97}. In
the case of multiply-imaged SNe the determination of the
time delay will be completely straightforward, as the time
of peak luminosity of a given SN will be determined to an
accuracy of better than a day as an integral part of the
SNAP observing program.  It is also often difficult to
determine the flux ratio of the images with good accuracy
(due to time variability of the source, extinction, etc.;
see~\citet{ms98}). As SNe light curves have been extensively
characterized, the flux ratios will be well determined.

It is often the modeling of the lens that limits the
precision of the determination of
$H_0$~\citep{keeton00,cohn01}.  A major impediment to
constraining lens profiles is the limited number of
constraints a lens system provides (at best, a handful of
image separations, relative brightnesses, and time
delays). Attempts to increase the number of constraints
generally rely on utilizing lensed images of the extended
background galaxy (arcs, Einstein rings, etc.;
see~\citet{keeton00,kkm01}). For quasars these attempts can
be hindered by the large brightness contrast between the
quasar source and the rest of the host galaxy. In the case
of SNe, however, the images conveniently turn themselves
off, allowing an uncontaminated view of the host. To
establish a baseline for the light ``contamination'' to the
SN curve from the host galaxies, SNAP will make deep images
of the hosts both before the SNe appear and after they fade
away. Subsequent very deep observations with a variety of
different instruments (e.g. NGST, CELT+AO) at a range of
wavelengths could also be made.  These images of the {\em
lensed}\/ host galaxy would provide important additional
constraints to the lens model. It
is again to be emphasized that the measure of $H_0$ with
time delays is independent of the standard-candleness of the
SNe, and is thus completely independent of
Cepheid-based measurements of the Hubble constant with type
Ia SNe~\citep{jha99}.

\section{Breaking the mass sheet degeneracy}
\label{S:mass_sheet}

Although the science in the preceding sections does not
depend on the SNe being standard candles, this property can
have important consequences for lensing studies. A drawback
of many lensing analyses done to date is that they suffer
from a mass-sheet degeneracy: a uniform sheet of matter
anywhere between the source and observer will generally
remain undetected. Because type Ia (and to a lesser degree
type II) SNe are standard candles, the {\em absolute}\/
magnification can be determined, yielding an additional
constraint (by changing the relative fluxes of all images
into absolute fluxes), and breaking this
degeneracy~\citep{kb98}.

An additional feature of SNAP is that, in the course of
returning to the same fields week after week looking for new
SNe, it will make very high quality weak lensing
maps. Coupling information from both weak and strong lensing
studies of a given field may yield additional insights---for
example, a mass concentration which causes a
large-separation multiple-imaging event may also be
detectable in the weak lensing map of the same field. Any
SNe (strongly lensed or not) superposed on a weak lensing
map will (in principle) allow for a breaking of the mass
sheet degeneracy.

\section{Conclusions}

The observation of multiply-imaged SNe would afford
independent measurements of cosmological
parameters. Although current high-$z$ SN surveys are
unlikely to observe a multiply-imaged SN system, SNAP would
be expected to see at least eight such events per year. These
systems would be a boon to science, independent of and in
addition to the primary goals of SNAP.

\acknowledgments

I thank Doug Eardley, Chris Fryer, Alex Kim, Saul Perlmutter, and
Jennie Traschen for illuminating discussions in the course of
this work. I am particularly grateful to Eric Agol, Greg
Aldering, Lars Bildsten, Leon Koopmans, and Warner Miller
for invaluable comments on the manuscript. This work was
partially supported by the NSF under grant PHY99-07949 to
the ITP.

\end{document}